\documentclass[letterpaper,journal]{IEEEtran}
\usepackage{amsmath,amsfonts,amssymb}
\usepackage{bm}
\usepackage{array}
\usepackage{booktabs}
\usepackage{multirow}
\usepackage{graphicx}
\usepackage{algorithmic}
\usepackage{algorithm}
\usepackage[caption=false,font=footnotesize]{subfig} 
\usepackage{xcolor}
\usepackage{cite}
\usepackage[
    colorlinks=true,
    linkcolor=blue,
    citecolor=blue,
    urlcolor=blue
]{hyperref}
\usepackage{orcidlink}

\makeatletter
\renewcommand{\@cite}[2]{%
  \textcolor{blue}{[#1\if@tempswa, #2\fi]}%
}
\makeatother
\usepackage{textcomp}
\usepackage{stfloats}
\usepackage{placeins}
\usepackage{url}
\usepackage{verbatim}
\usepackage{balance}


\begin{document}

\title{Compound Interference Recognition for LR-FHSS Satellite IoT Uplinks via Multi-Domain Instance Fusion}

\author{Haitao Xu\orcidlink{0009-0009-7667-0522}, Boxiang He\orcidlink{0000-0002-9235-1144}, Shilian Wang\orcidlink{0000-0003-4132-8750}, and Yinan Jiang\orcidlink{0009-0000-9355-3680}%
\thanks{H. Xu, B. He, S. Wang, and Y. Jiang are with the College of Electronic Science and Technology, National University of Defense Technology, Changsha 410003, P. R. China (e-mail: HaitaoXu1997@outlook.com; boxianghe1@bjtu.edu.cn; wangsl@nudt.edu.cn; jiangyinan778@168.com).}%
\thanks{This work was supported in part by the National Natural Science Foundation of China under Grant 62501612, and in part by the China Postdoctoral Science Foundation under Grant 2025M774418.}}


\maketitle

\begin{abstract}
Long range-frequency hopping spread spectrum (LR-FHSS) is a promising uplink physical layer for massive low Earth orbit satellite Internet of Things, where low power terminals report short packets from wide area regions with limited terrestrial infrastructure.
However, satellite IoT links are exposed to external interference, and the coexistence of multiple interference components can severely degrade receiver reliability and complicate interference mitigation.
Existing recognition methods either focus on single interference scenarios or treat each compound interference combination as an independent class, leading to limited generalization or poor scalability.
To address this problem, this paper formulates LR-FHSS uplink compound interference recognition as a multi-instance multi-label learning problem and proposes a multi-domain instance fusion method.
The proposed method fuses local instances from the time-frequency and frequency domains and aggregates their predictions for bag-level multi-label recognition.
A dataset construction pipeline is developed based on the US915 LR-FHSS configuration and incorporates shadowed-Rician fading and time-varying Doppler to emulate practical satellite communication conditions.
Considering the difficulty of obtaining labeled compound interference samples in practice, single-to-compound generalization and few-shot compound interference adaptation are investigated as two practical receiver deployment scenarios.
Experimental results show that the proposed method improves the overall exact accuracy over the strongest baseline by 14.71 percentage points in single-to-compound generalization and by 14.81 percentage points in few-shot compound interference adaptation for $r=1$.
\end{abstract}


\section{Introduction}
\IEEEPARstart{S}{atellite} Internet of Things (IoT) extends terrestrial IoT services to remote and infrastructure-limited regions where continuous terrestrial coverage is technically difficult or economically infeasible \cite{sat-iot-trends}. 
Low Earth orbit (LEO) satellites are particularly attractive for direct-to-satellite IoT because their lower altitude reduces link distance and propagation delay compared with higher orbit systems, while broad satellite coverage enables low power terminals to transmit short packets without nearby terrestrial gateways \cite{noma-aloha-leo-iot,lora-leo-sg}. 
These properties make LEO satellite IoT suitable for low duty cycle sensing applications that require wide area coverage, including agriculture, environmental monitoring, maritime monitoring, logistics tracking, and utility metering \cite{sat-iot-survey,dts-iot-collision}.

Long range wide area network (LoRaWAN) is a practical candidate for LEO direct-to-satellite IoT with low power terminals, especially for uplink-dominated sensing applications.
LoRaWAN regional parameters specify long range-frequency hopping spread spectrum (LR-FHSS) as an uplink physical layer waveform.
An LR-FHSS packet consists of repeated physical layer headers and coded payload fragments transmitted over pseudorandom hopping carriers, thereby improving signal coexistence and robustness against narrowband interference in dense IoT deployments \cite{lrfhss-overview,dts-lrfhss-simulation}.
Existing LR-FHSS studies mainly analyze packet collisions, outage under satellite channels, header replication and recovery, and resource allocation for header and payload transmissions \cite{lrfhss-outage,lrfhss-acrda,lrfhss-nc-header,lrfhss-orthogonal}.
Although these studies characterize access performance and packet reliability in LR-FHSS satellite IoT, receiver side recognition of external interference from received signal samples remains insufficiently studied.

External interference is a serious reliability and security risk for satellite IoT links.
In integrated satellite and terrestrial IoT networks, external interference may originate from other satellite systems or terrestrial sources and directly affect outage performance \cite{sat-iot-ostn-interference}.
Exposed links between satellites and ground terminals make satellite IoT vulnerable to malicious or intelligent interference attacks, which can reduce transmission rate, degrade throughput, or even interrupt network operation \cite{siot-beam-antijamming,sat-iot-antijamming}.
This risk becomes more pronounced when multiple interference components coexist or when intelligent interference reacts to LR-FHSS transmissions in time-varying local regions.
Such conditions require compound interference recognition to identify all active components for interference mitigation.
In practice, labeled compound interference samples are difficult to obtain during initial receiver development, because compound interference events are sporadic and their component labels require additional monitoring or offline analysis.
Therefore, an important practical problem is whether a model trained only with non-interference and single interference samples can generalize to compound interference encountered during operation.
When a small number of labeled compound interference samples become available, receiver fine-tuning is also needed to improve recognition performance.

Most existing recognition methods focus on single interference scenarios, where convolutional neural network (CNN)-based, transformer-based, multi-domain, and multimodal models are trained to map received signal samples to a single interference category \cite{wireless-interference-cnn,wir-transformer,wireless-interference-multidomain,wireless-interference-multimodal,m2-net}.
For compound interference recognition, existing studies mainly adopt two formulations.
One is single-label classification, which treats each compound interference combination as an independent class \cite{radar-compound-jamming-dual-channel,fh-compound-interference-classification}. 
This formulation can identify known combinations, but its predefined class space grows exponentially with the number of basic interference categories, reaching $2^C-1$ output nodes for $C$ categories. 
The other is multi-label classification, which represents active components with a multi-hot label vector and predicts this vector from received signal samples \cite{compound-signal-mlamc,radar-compound-jamming-cvcnn,compound-interference-uav}. 
This formulation avoids enumerating interference combinations and is therefore more scalable.
However, existing multi-label models usually rely on a single global representation, which may weaken localized interference evidence that appears within only a subset of LR-FHSS hopping fragments, time slots, and frequency bins.
This limitation is particularly relevant to reactive interference, whose evidence is time-varying and locally coupled with LR-FHSS transmissions.

\begin{figure}[!t]
\centering
\includegraphics[width=\linewidth]{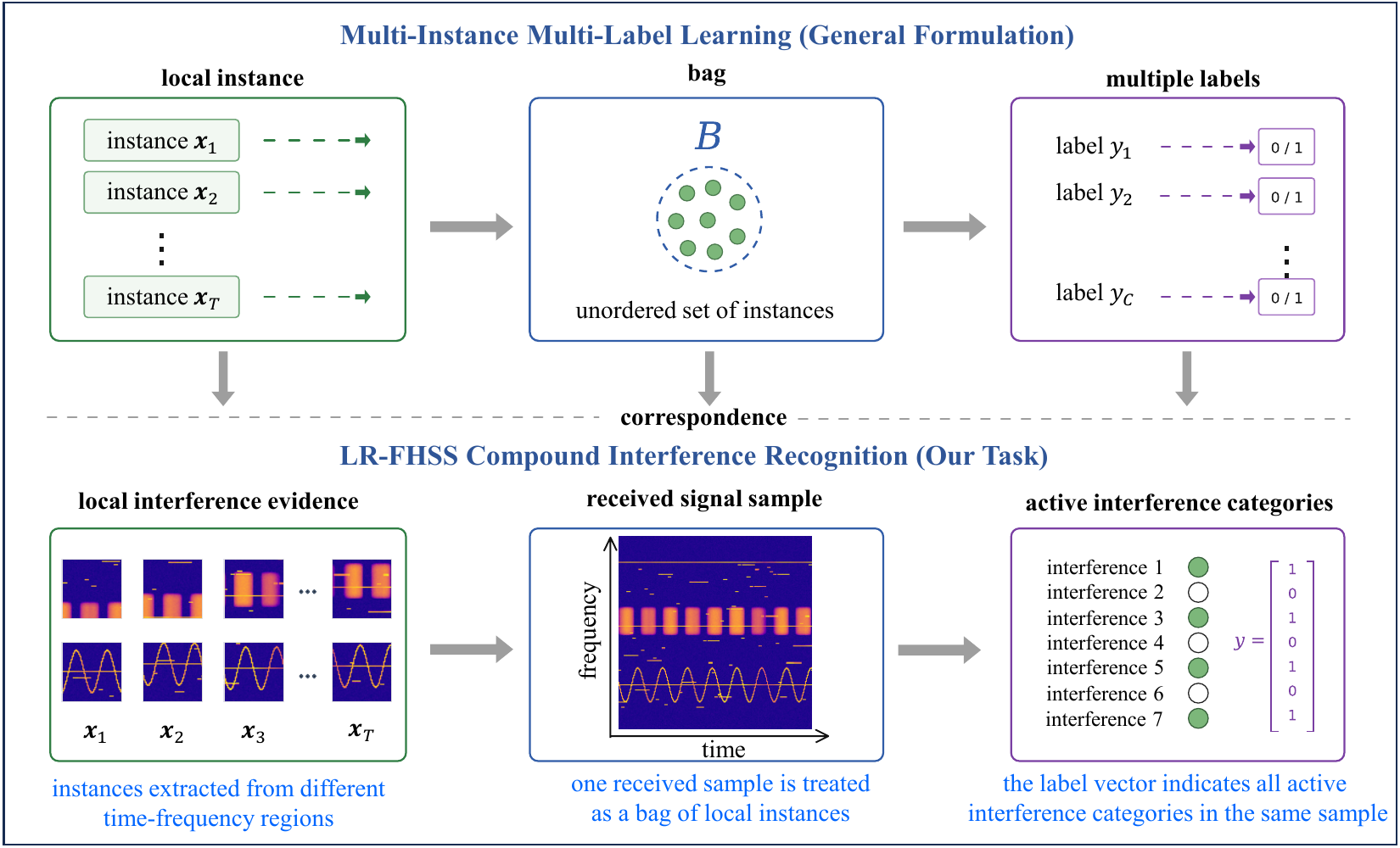}
\caption{Correspondence between multi-instance multi-label learning and LR-FHSS compound interference recognition.}
\label{fig:miml_mapping}
\vspace{-0em}
\end{figure}

Multi-instance multi-label learning represents each sample as a bag of local instances and supervises the bag with a multi-hot label vector \cite{miml-foundation}.
As illustrated in Fig.~\ref{fig:miml_mapping}, this formulation matches LR-FHSS compound interference recognition because one received signal sample can be represented as a bag of local instances, where different local regions may contain evidence of multiple interference categories.
The bag-level multi-hot label indicates which interference categories are present in the whole sample, while instance-level labels for individual local regions are unavailable.
Therefore, the multi-instance multi-label formulation allows localized interference evidence to be exploited without requiring annotations for individual time-frequency patches or frequency segments.
Studies in \cite{overlapping-lpi-miml,miml-gan} have applied this formulation to overlapping signal waveform recognition by treating each time-frequency representation as a bag and its local regions as instances, demonstrating its potential for compound interference recognition.
Meanwhile, DINOv3 is trained with large-scale self-supervised learning on LVD-1689M and produces dense local representations suitable for extracting local instances from time-frequency signal representations \cite{vit,dinov3}.

Motivated by these observations, we develop a multi-domain instance fusion method for LR-FHSS uplink compound interference recognition under a multi-instance multi-label learning formulation. 
The key idea is to fuse local instances from the time-frequency and frequency domains, allowing active interference components to be identified from localized evidence rather than from a single global representation.
The main contributions of this paper are summarized as follows.

\begin{itemize}
\item We formulate LR-FHSS uplink compound interference recognition as a multi-instance multi-label learning problem.
In this formulation, each received sample is represented as a bag of local instances and is supervised only by a bag-level multi-hot label vector.
This formulation avoids enumerating compound interference combinations as independent classes, matches the localized evidence structure of compound interference, and does not require instance-level annotations.

\item We propose a multi-domain instance fusion method for compound interference recognition.
The time-frequency branch uses a frozen pretrained DINOv3 ViT backbone to extract local instances from time-frequency images, while the frequency branch uses a lightweight auxiliary network to extract local instances from frequency sequences.
The local instances from the two domains are projected into a common feature space, and max pooling selects the strongest local evidence for each interference category to obtain bag-level multi-label predictions.
This design exploits complementary localized interference evidence from the time-frequency and frequency domains while keeping the pretrained DINOv3 ViT backbone frozen and optimizing only task-specific modules.

\item We develop a dataset construction pipeline for LR-FHSS uplink compound interference recognition that includes conventional interference categories and two challenging reactive interference categories.
The dataset uses LR-FHSS as the background communication signal, follows the US915 LR-FHSS configuration, and incorporates shadowed-Rician fading and time-varying Doppler to emulate practical satellite communication conditions.
Based on this dataset, we evaluate the proposed method under two practical receiver deployment scenarios.
The single-to-compound generalization scenario considers the case where only non-interference and single interference samples are available during training, whereas the few-shot compound interference adaptation scenario considers receiver fine-tuning with a small number of labeled compound interference samples.
Extensive experiments demonstrate the effectiveness of the proposed method, and further analyses reveal the recognition difficulty of reactive interference and the role of each signal domain.
\end{itemize}

The remainder of this paper is organized as follows. 
Section II presents the system model and problem formulation. 
Section III describes the proposed multi-domain instance fusion method for LR-FHSS uplink compound interference recognition.
Section IV reports the experimental results and analysis.
Section V concludes this paper.

\emph{Notation:} Throughout this paper, scalars, vectors, and matrices are denoted by italic lowercase letters $x$, bold italic lowercase letters $\bm{x}$, and bold italic uppercase letters $\bm{X}$, respectively. Calligraphic letters, e.g., $\mathcal{C}$, denote sets. $\mathbb{R}$ and $\mathbb{C}$ denote the sets of real and complex numbers, respectively. The operator $[\cdot]^{\mathsf{T}}$ denotes transpose. The symbols $\ast$ and $\cdot$ denote convolution and multiplication, respectively. The operator $|\cdot|$ denotes set cardinality when applied to a set. The indicator function $\mathbb{I}(\cdot)$ returns 1 if its argument is true and 0 otherwise. $j=\sqrt{-1}$ is the imaginary unit.

\begin{figure*}[!t]
\centering
\includegraphics[width=0.90\textwidth]{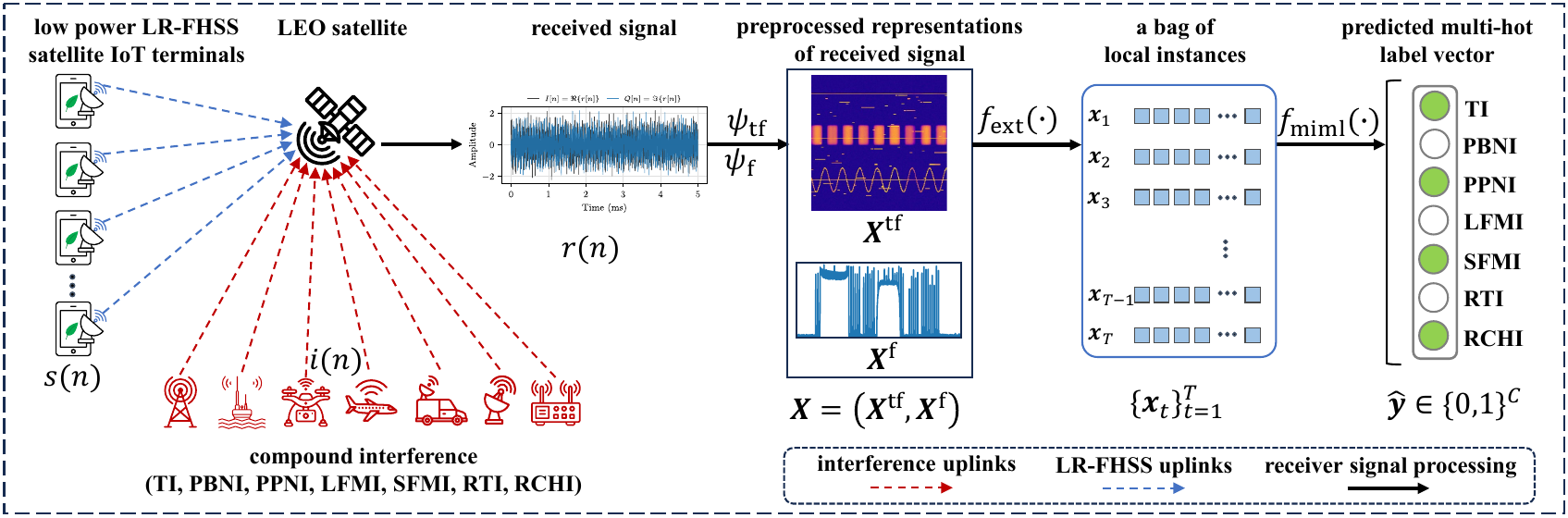}
\caption{System model and receiver processing flow for LR-FHSS uplink compound interference recognition.}
\label{fig:system_model}
\vspace{-0em}
\end{figure*}

\section{System Model and Problem Formulation}
\label{sec:system_model}
This section presents the system model for LR-FHSS satellite IoT uplink compound interference recognition and formulates the recognition task as a multi-instance multi-label learning problem.

\subsection{System Model}
As illustrated in Fig.~\ref{fig:system_model}, this paper considers a direct LR-FHSS uplink from low power LoRaWAN terminals to an LEO satellite receiver. At the satellite receiver, the received signal may contain the desired LR-FHSS signal and multiple coexisting interference components. 
The receiver computes a short-time Fourier transform (STFT) power spectrogram from the received signal and constructs a time-frequency image and a frequency sequence for interference recognition.
These two representations are then converted into a bag of local instances, and the active interference categories are represented by a multi-hot label vector. Thus, LR-FHSS compound interference recognition is formulated as a multi-instance multi-label learning problem.

Let $\mathcal{C}$ denote the interference category set, and let $C=|\mathcal{C}|$ be the number of interference categories.
For a sample with label vector $\bm{y}=[y_1,\ldots,y_C]^{\mathsf{T}}\in\{0,1\}^{C}$, the received complex baseband signal at the satellite is modeled as
\begin{equation}
\begin{aligned}
r(n) &= \sqrt{P_{\mathrm{s}}}\,h_0(n)\bar{s}(n)\mathrm{e}^{j(\phi_{D,0}(n)+\phi_0)} \\
&\quad + \sum_{c=1}^{C} y_c \sqrt{P_{\mathrm{ref}}}\,h_c(n)\bar{i}_c(n)
\mathrm{e}^{j(\phi_{D,c}(n)+\phi_c)} + w(n),
\end{aligned}
\label{eq:received_signal}
\end{equation}
where $\bar{s}(n)\in\mathbb{C}$ is the normalized LR-FHSS baseband signal, $\bar{i}_c(n)\in\mathbb{C}$ is the normalized $c$-th interference component, $P_{\mathrm{s}}$ is the LR-FHSS background signal power, and $P_{\mathrm{ref}}$ is the reference power of a single interference component.
The binary variable $y_c$ indicates whether the $c$-th interference category is active.
The coefficients $h_0(n)$ and $h_c(n)$ denote shadowed-Rician fading coefficients, $\phi_0$ and $\phi_c$ denote phase offsets, and $w(n)$ is complex additive white Gaussian noise (AWGN) with variance $\sigma_w^2$.
The Doppler phase of source $\ell\in\{0,1,\ldots,C\}$ is
\begin{equation}
\phi_{D,\ell}(n)=2\pi\sum_{q=0}^{n}\frac{f_{\mathrm{dop},\ell}(q)}{f_s},
\end{equation}
where $f_{\mathrm{dop},\ell}(q)$ is the time-varying Doppler shift and $f_s$ is the sampling rate.
The normalized waveforms are obtained as
\begin{equation}
\bar{s}(n)=\frac{s(n)}
{\sqrt{\frac{1}{L}\sum_{q=0}^{L-1}|s(q)|^2}},
\end{equation}
\begin{equation}
\bar{i}_c(n)=\frac{i_c(n)}
{\sqrt{\frac{1}{L}\sum_{q=0}^{L-1}|i_c(q)|^2}},
\end{equation}
where $s(n)$ and $i_c(n)$ are the corresponding waveforms before power normalization, and $L$ is the length of each received signal sample.
Each active interference component is scaled to the same reference power $P_{\mathrm{ref}}$, which provides a common power reference for defining interference-to-signal ratio (ISR) and interference-to-noise ratio (INR).
The ISR and INR are then defined as
\begin{equation}
\mathrm{ISR}=10\log_{10}\frac{P_{\mathrm{ref}}}{P_{\mathrm{s}}},
\end{equation}
\begin{equation}
\mathrm{INR}=10\log_{10}\frac{P_{\mathrm{ref}}}{\sigma_w^2}.
\end{equation}
Thus, $P_{\mathrm{s}}=P_{\mathrm{ref}}/10^{\mathrm{ISR}/10}$ and $\sigma_w^2=P_{\mathrm{ref}}/10^{\mathrm{INR}/10}$.

An LR-FHSS complex baseband waveform before power normalization is modeled as
\begin{equation}
s(n)=a(n)\mathrm{e}^{j2\pi\sum_{q=0}^{n}\frac{f_{\mathrm{FHSS}}(q)}{f_s}},
\end{equation}
where $a(n)$ denotes the coded and Gaussian minimum shift keying (GMSK) modulated LR-FHSS baseband sequence before hopping, and $f_{\mathrm{FHSS}}(n)$ denotes the LR-FHSS hopping frequency, which remains constant within each hop.

\begin{figure}[!t]
\centering
\subfloat[RTI]{%
\includegraphics[width=0.48\columnwidth]{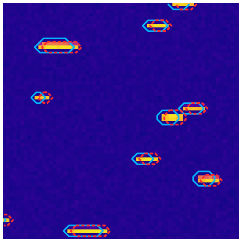}}
\hfill
\subfloat[RCHI]{%
\includegraphics[width=0.48\columnwidth]{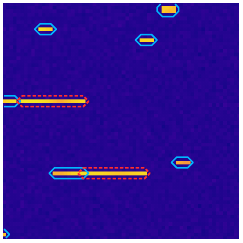}}
\caption{Local time-frequency relationships between the LR-FHSS hopping signal and reactive interference components. Blue and red contours mark the LR-FHSS hopping signal and the reactive interference, respectively.}
\label{fig:reactive_interference}
\vspace{-0em}
\end{figure}

The interference components considered in this paper include tone interference (TI), partial band noise interference (PBNI), periodic pulsed noise interference (PPNI), linear frequency modulated interference (LFMI), sinusoidal frequency modulated interference (SFMI) \cite{fh-compound-interference-classification,fh-multinode-interference-recognition}, reactive tracking interference (RTI), and reactive capture-and-hold interference (RCHI) \cite{im-fhss-unified-antijamming,fhma-follower-jamming}. 
As illustrated in Fig.~\ref{fig:reactive_interference}, RTI and RCHI are challenging to recognize because they adapt to the target hopping frequency.
The desired LR-FHSS waveform $s(n)$ is treated as the background communication signal, and the baseband waveform of each interference component $i_c(n)$ is defined below.

\emph{1) TI:} One or more complex tones are generated at fixed frequency offsets. The complex baseband waveform is modeled as
\begin{equation}
i_{\mathrm{TI}}(n)=\sum_{u=1}^{U}
\mathrm{e}^{j\left(2\pi f_{u}\frac{n}{f_s}+\phi_{u}\right)},
\end{equation}
where $U$ is the number of tones; $f_u$ and $\phi_u$ are the frequency offset and initial phase of the $u$-th tone, respectively.

\emph{2) PBNI:} An AWGN sequence is filtered over a partial band and shifted to a center frequency. The complex baseband waveform is modeled as
\begin{equation}
i_{\mathrm{PBNI}}(n)=\bigl(g_B \ast \nu\bigr)(n)
\cdot \mathrm{e}^{j\left(2\pi f_{\mathrm{cen}}\frac{n}{f_s}+\phi\right)},
\end{equation}
where $\nu(n)$ is AWGN; $g_B(n)$ is a band-limiting filter with bandwidth $B$; $f_{\mathrm{cen}}$ and $\phi$ denote the center frequency and initial phase, respectively.

\emph{3) PPNI:} An AWGN sequence is filtered over a partial band, shifted to a center frequency, and gated by a periodic pulse envelope. 
The complex baseband waveform is modeled as
\begin{equation}
i_{\mathrm{PPNI}}(n)=p(n)\cdot\bigl(g_B \ast \nu\bigr)(n)
\cdot \mathrm{e}^{j\left(2\pi f_{\mathrm{cen}}\frac{n}{f_s}+\phi\right)},
\end{equation}
where $p(n)$ denotes the periodic pulse envelope with pulse period $T_p$ and duty cycle $\tau_p$.

\emph{4) LFMI:} The instantaneous frequency is swept linearly over a specified bandwidth within each sweep period. 
The complex baseband waveform is modeled as
\begin{equation}
i_{\mathrm{LFMI}}(n)
=\mathrm{e}^{j\left(2\pi\sum_{q=0}^{n}\frac{f_{\mathrm{LFMI}}(q)}{f_s}+\phi\right)},
\end{equation}
where $f_{\mathrm{LFMI}}(n)=f_{\mathrm{cen}}-B_{\mathrm{LFMI}}/2+B_{\mathrm{LFMI}}((n+n_{\mathrm{LFMI},0})\allowbreak \bmod N_{\mathrm{LFMI}})/N_{\mathrm{LFMI}}$ is the instantaneous frequency, $B_{\mathrm{LFMI}}$ is the sweep bandwidth, $N_{\mathrm{LFMI}}$ is the sweep period in samples, and $n_{\mathrm{LFMI},0}$ is the sweep start instant.

\emph{5) SFMI:} The instantaneous frequency is varied sinusoidally around a center frequency within each modulation period. 
The complex baseband waveform is modeled as
\begin{equation}
i_{\mathrm{SFMI}}(n)
=\mathrm{e}^{j\left(2\pi\sum_{q=0}^{n}\frac{f_{\mathrm{SFMI}}(q)}{f_s}+\phi\right)},
\end{equation}
where $f_{\mathrm{SFMI}}(n)=f_{\mathrm{cen}}+\Delta f_{\mathrm{SFMI}}\sin(2\pi((n+n_{\mathrm{SFMI},0})\allowbreak \bmod N_{\mathrm{SFMI}})/N_{\mathrm{SFMI}})$ is the instantaneous frequency, $\Delta f_{\mathrm{SFMI}}$ is the frequency deviation, $N_{\mathrm{SFMI}}$ is the modulation period in samples, and $n_{\mathrm{SFMI},0}$ is the modulation start instant.

\emph{6) RTI:} A reactive interference signal follows the LR-FHSS hopping frequency after a response delay. 
The complex baseband waveform is modeled as
\begin{equation}
i_{\mathrm{RTI}}(n)=
\begin{cases}
0, & n<\kappa_0+N_d,\\
\mathrm{e}^{j\left(2\pi\sum_{q=\kappa_0+N_d}^{n}
\frac{f_{\mathrm{FHSS}}(q-N_d)}{f_s}+\phi\right)}, & n\ge \kappa_0+N_d,
\end{cases}
\end{equation}
where $\kappa_0$ is the initial capture instant, $N_d=\operatorname{round}(d f_s)$ is the response delay in samples, and $d$ is the delay in seconds.

\emph{7) RCHI:} A reactive interference signal periodically captures the LR-FHSS hopping frequency after a response delay and holds it until the next capture instant. 
The complex baseband waveform is modeled as
\begin{equation}
i_{\mathrm{RCHI}}(n)=
\begin{cases}
0, & n<\kappa_0+N_d,\\
\mathrm{e}^{j\left(2\pi\sum_{q=\kappa_0+N_d}^{n}
\frac{f_{\mathrm{FHSS}}(\kappa(q))}{f_s}+\phi\right)}, & n\ge \kappa_0+N_d,
\end{cases}
\end{equation}
where
\begin{equation}
\kappa(q)=\max\{\kappa_m:\kappa_m+N_d\le q\},
\end{equation}
$\kappa_m=\kappa_0+mN_{\mathrm{cap}}$ denotes the $m$-th capture instant, and $N_{\mathrm{cap}}$ is the capture period in samples.

\subsection{Problem Formulation}
The receiver first maps each received signal to an STFT power spectrogram. Let $\bm{S}\in\mathbb{C}^{F\times M}$ denote the STFT matrix, where $F$ and $M$ are the numbers of frequency bins and time frames, respectively. 
The STFT matrix is computed as
\begin{equation}
S(k,m)=\sum_{n} r(n)g(n-mR)\mathrm{e}^{-j\frac{2\pi kn}{N_{\mathrm{fft}}}},
\end{equation}
where $g(\cdot)$ is the analysis window, $R$ is the STFT hop size, and $N_{\mathrm{fft}}$ is the fast Fourier transform (FFT) size.
The power spectrogram $\bm{P}\in\mathbb{R}^{F\times M}$ is obtained as $P(k,m)=|S(k,m)|^2$. 
From $\bm{P}$, we construct a time-frequency image and a frequency sequence as
\begin{equation}
\bm{X}^{\mathrm{tf}}
=
\psi_{\mathrm{tf}}(\bm{P})
\in\mathbb{R}^{3\times H\times W},
\end{equation}
\begin{equation}
\bm{X}^{\mathrm{f}}
=
\psi_{\mathrm{f}}(\bm{P})
\in\mathbb{R}^{F},
\end{equation}
where $H$ and $W$ are the image height and width, respectively;
$\psi_{\mathrm{tf}}(\cdot)$ maps the power spectrogram to the RGB time-frequency image;
$\psi_{\mathrm{f}}(\cdot)$ maps the power spectrogram to the frequency sequence.
Thus, the input sample of the recognition model is represented as
$\bm{X}=(\bm{X}^{\mathrm{tf}},\bm{X}^{\mathrm{f}})$.

\begin{figure}[!t]
\centering
\includegraphics[width=\linewidth]{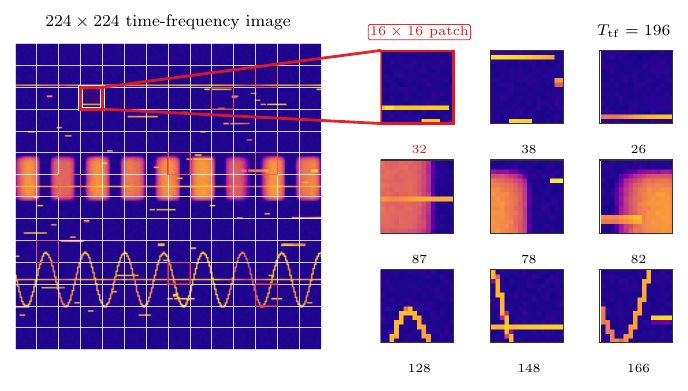}
\caption{Illustration of local instance construction in the time-frequency branch. A $224\times224$ time-frequency image is partitioned into $16\times16$ patches, and the highlighted patch is enlarged on the right. The patch embeddings extracted from these local patches are treated as local instances, resulting in $T_{\mathrm{tf}}=196$ time-frequency instances.}
\label{fig:local_instance_view}
\vspace{-0em}
\end{figure}

The recognition objective is to predict all active interference categories from the multi-domain input $\bm{X}$. 
Under the multi-instance multi-label formulation, each input sample is converted into a bag of local instances. 
Here, an instance refers to a local representation extracted from the time-frequency image or the frequency sequence.
For example, Fig.~\ref{fig:local_instance_view} illustrates that a time-frequency image is partitioned into fixed-size patches, and these patches are encoded by the instance extractor into local instances.
The label vector specifies which interference categories are present in the whole sample, but does not indicate which local time-frequency patch or frequency segment supports each category.
Therefore, only bag-level supervision is available.

Let $f_{\mathrm{ext}}(\cdot)$ denote the instance extraction function that maps an input sample to local instances. Given $\bm{X}$, the resulting instance bag is written as
\begin{equation}
\label{eq:instance_extraction}
\left[\bm{x}_{1},\ldots,\bm{x}_{T}\right]^{\mathsf{T}}=f_{\mathrm{ext}}(\bm{X})\in\mathbb{R}^{T\times D_{\mathrm{o}}},
\end{equation}
where $\bm{x}_{t}$ denotes the $t$-th local instance, $T$ is the number of instances in the bag, and $D_{\mathrm{o}}$ is the common instance dimension.
Let $f_{\mathrm{miml}}(\cdot)$ denote the multi-instance multi-label classification function that maps a bag of instances to a multi-label prediction. The predicted probability vector is given by 
\begin{equation}
\label{eq:miml_prediction}
\bm{p}=[p_1,\ldots,p_C]^{\mathsf{T}}=f_{\mathrm{miml}}\left(f_{\mathrm{ext}}(\bm{X})\right)\in[0,1]^C,
\end{equation}
where $p_c$ denotes the predicted probability of the $c$-th interference category. 
The predicted multi-hot label vector $\hat{\bm{y}}\in\{0,1\}^C$ is obtained by applying a category-specific threshold to $\bm{p}$, i.e., $\hat{y}_c=\mathbb{I}\left(p_c>\tau_c\right)$, where $\tau_c$ is the threshold for the $c$-th category.

Let $\mathcal{D}=\left\{(\bm{X}_i,\bm{y}_i)\right\}_{i=1}^{N_{\mathrm{tr}}}$ denote the training dataset with $N_{\mathrm{tr}}$ samples, where $\bm{X}_i$ is the $i$-th input sample and $\bm{y}_i$ is the corresponding ground truth multi-hot label vector.
The LR-FHSS compound interference recognition problem is to learn the instance extraction function $f_{\mathrm{ext}}(\cdot)$ and the multi-instance multi-label classification function $f_{\mathrm{miml}}(\cdot)$ by minimizing the empirical multi-label loss
\begin{equation}
\begin{aligned}
\left(f_{\mathrm{ext}}^{\star},f_{\mathrm{miml}}^{\star}\right)
&=
\operatorname*{arg\,min}_{f_{\mathrm{ext}},f_{\mathrm{miml}}}
\frac{1}{N_{\mathrm{tr}}}
\sum_{i=1}^{N_{\mathrm{tr}}}
\mathcal{L}\left(
f_{\mathrm{miml}}\left(f_{\mathrm{ext}}(\bm{X}_i)\right),\bm{y}_i
\right),
\end{aligned}
\end{equation}
where $f_{\mathrm{ext}}^{\star}$ and $f_{\mathrm{miml}}^{\star}$ denote the corresponding optimal functions, and $\mathcal{L}(\cdot,\cdot)$ denotes a multi-label loss function.
The concrete construction of $f_{\mathrm{ext}}(\cdot)$ and $f_{\mathrm{miml}}(\cdot)$ is described in the next section.

\section{Proposed Multi-Domain Instance Fusion Method}
\label{sec:method}

\begin{figure*}[!t]
\centering
\includegraphics[width=0.90\textwidth]{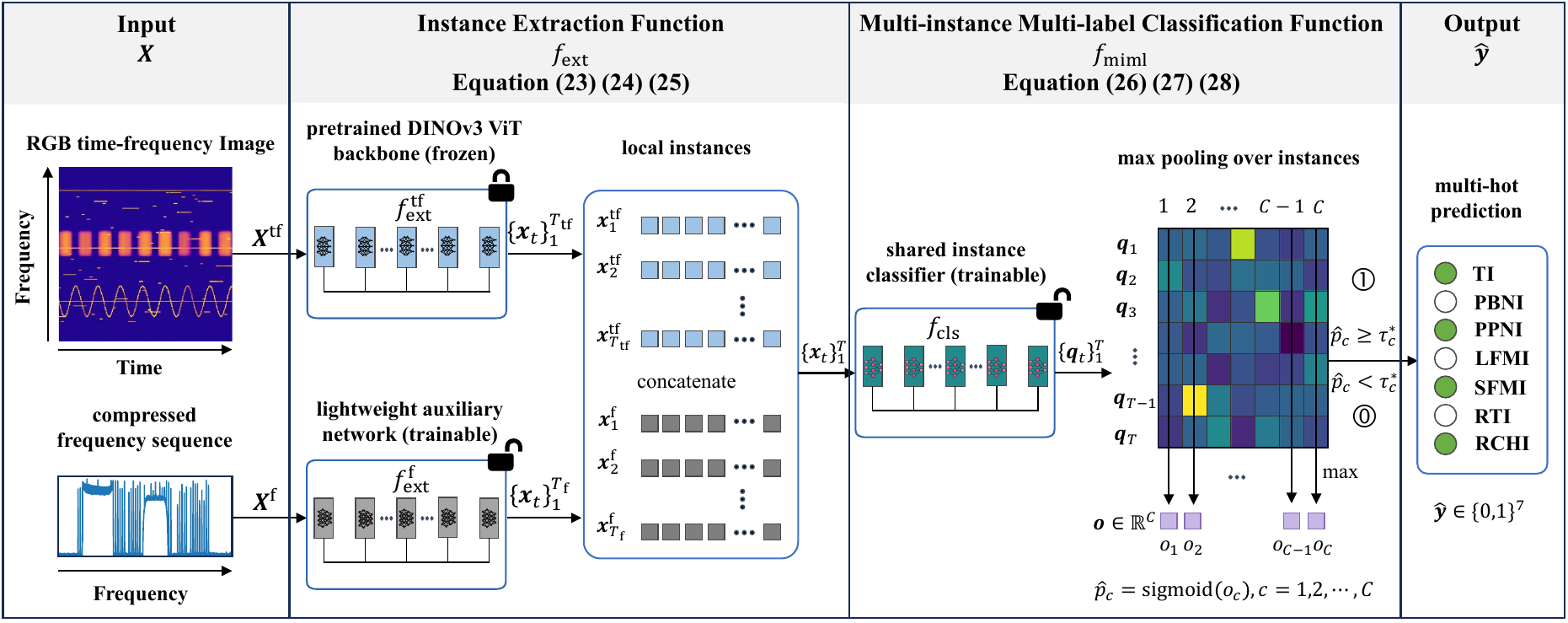}
\caption{Proposed multi-domain instance fusion method.}
\label{fig:architecture}
\vspace{-0em}
\end{figure*}

This section presents the proposed multi-domain instance fusion method, which instantiates the instance extraction function $f_{\mathrm{ext}}(\cdot)$ in Section~\ref{sec:system_model} with a time-frequency extractor $f_{\mathrm{ext}}^{\mathrm{tf}}(\cdot)$ and a frequency extractor $f_{\mathrm{ext}}^{\mathrm{f}}(\cdot)$.
As shown in Fig.~\ref{fig:architecture}, the method extracts time-frequency instances from the time-frequency image using a frozen pretrained DINOv3 ViT backbone and extracts frequency instances from the frequency sequence using a lightweight auxiliary branch.
The extracted instances from the two domains are projected to a common feature space, concatenated into a unified instance bag, classified by a shared instance classifier, and aggregated by max pooling to obtain bag-level multi-label predictions.
With the DINOv3 ViT backbone frozen, only the projection layers, the frequency branch, and the shared instance classifier are optimized during training.

\subsection{Multi-Domain Instance Extraction and Fusion}
\label{subsec:instance_extraction}

\subsubsection{Time-Frequency Instance Extraction}
The time-frequency extractor $f_{\mathrm{ext}}^{\mathrm{tf}}(\cdot)$ maps the time-frequency image $\bm{X}^{\mathrm{tf}}\in\mathbb{R}^{3\times H\times W}$ to local time-frequency instances. 
A frozen pretrained DINOv3 ViT backbone is used to extract local time-frequency instances from the time-frequency image \cite{vit,dinov3}.
Let $f_{\mathrm{DINO}}(\cdot)$ denote the frozen DINOv3 ViT instance extractor. 
Let $\pi_{\mathrm{tf}}(\cdot)$ denote a trainable projection layer to align the DINOv3 instance dimension to the common instance dimension $D_{\mathrm{o}}$.
The time-frequency instances are obtained as
\begin{equation}
\begin{aligned}
\relax\left[\bm{x}^{\mathrm{tf}}_1,\ldots,\bm{x}^{\mathrm{tf}}_{T_{\mathrm{tf}}}\right]^{\mathsf{T}}
&=f_{\mathrm{ext}}^{\mathrm{tf}}(\bm{X}^{\mathrm{tf}}) \\
&=\pi_{\mathrm{tf}}\left(\operatorname{Patch}\left(f_{\mathrm{DINO}}(\bm{X}^{\mathrm{tf}})\right)\right)
\in\mathbb{R}^{T_{\mathrm{tf}}\times D_{\mathrm{o}}},
\end{aligned}
\end{equation}
where $\operatorname{Patch}(\cdot)$ removes the class token and register tokens from the DINOv3 ViT output sequence and keeps the remaining patch tokens.
In the implementation, $H=W=224$, the ViT patch size is $16$, and therefore $T_{\mathrm{tf}}=196$; $D_{\mathrm{o}}=768$.

\subsubsection{Frequency Instance Extraction}
The frequency extractor $f_{\mathrm{ext}}^{\mathrm{f}}(\cdot)$ maps the frequency sequence $\bm{X}^{\mathrm{f}}\in\mathbb{R}^{F}$ to local frequency instances. 
Let $f_{\mathrm{aux}}(\cdot)$ denote the lightweight auxiliary branch for frequency instance extraction.
Let $\pi_{\mathrm{f}}(\cdot)$ denote the projection layer to align the auxiliary branch output dimension to the common instance dimension $D_{\mathrm{o}}$.
The frequency instances are obtained as
\begin{equation}
\begin{aligned}
\relax\left[\bm{x}^{\mathrm{f}}_1,\ldots,\bm{x}^{\mathrm{f}}_{T_{\mathrm{f}}}\right]^{\mathsf{T}}
&=f_{\mathrm{ext}}^{\mathrm{f}}\left(\bm{X}^{\mathrm{f}}\right)\\
&=\pi_{\mathrm{f}}\left(f_{\mathrm{aux}}\left(\bm{X}^{\mathrm{f}}\right)\right)
\in\mathbb{R}^{T_{\mathrm{f}}\times D_{\mathrm{o}}}.
\end{aligned}
\end{equation}
In the implementation, the frequency branch outputs $T_{\mathrm{f}}=8$ frequency instances.

By concatenating the projected instances from the two extractors, the unified instance bag in \eqref{eq:instance_extraction} is obtained as
\begin{equation}
\begin{aligned}
\left[\bm{x}_{1},\ldots,\bm{x}_{T}\right]^{\mathsf{T}}
&=\operatorname{Concat}\left(
 f_{\mathrm{ext}}^{\mathrm{tf}}(\bm{X}^{\mathrm{tf}}),
 f_{\mathrm{ext}}^{\mathrm{f}}(\bm{X}^{\mathrm{f}})
 \right) \in\mathbb{R}^{T\times D_{\mathrm{o}}},
\end{aligned}
\end{equation}
where $T=T_{\mathrm{tf}}+T_{\mathrm{f}}$.

\subsection{Bag-Level Multi-Label Prediction}
\label{subsec:instance_fusion_prediction}
After obtaining the unified instance bag from $f_{\mathrm{ext}}(\cdot)$, the method applies a shared instance classifier to compute instance-level logits and aggregates them by max pooling.
This design follows the multi-instance assumption that a bag-level label can be activated by strong local evidence for the corresponding interference category.
For each local instance $\bm{x}_t$, the shared instance classifier $f_{\mathrm{cls}}(\cdot)$ produces an instance-level logit vector

\begin{equation}
\bm{q}_t=f_{\mathrm{cls}}(\bm{x}_t)\in\mathbb{R}^{C},\quad t=1,2,\ldots,T.
\end{equation}
The bag-level logit for the $c$-th interference category is then obtained by
\begin{equation}
o_c=\max_{1\le t\le T} q_{t,c},
\end{equation}
where $q_{t,c}$ is the logit of the $c$-th category at the $t$-th instance.
This max-pooling operation selects the strongest local evidence for each interference category, allowing the model to activate a category even when its evidence appears only in a limited portion of the received sample.

After max pooling, the bag-level logit vector is denoted by $\bm{o}=[o_1,\ldots,o_C]^{\mathsf{T}}$.
The bag-level probability vector in \eqref{eq:miml_prediction} is obtained as
\begin{equation}
\bm{p}=[p_1,\ldots,p_C]^{\mathsf{T}}
=\operatorname{sigmoid}(\bm{o})\in[0,1]^C.
\end{equation}

\subsection{Loss Function and Threshold Selection}
\label{subsec:loss_threshold}
\subsubsection{Asymmetric Loss}
To handle label imbalance in multi-label compound interference recognition, the model is trained with asymmetric loss (ASL) \cite{asymmetric-loss}.
In practical LR-FHSS monitoring, each received sample usually contains only a small subset of interference categories, while most categories are absent.
Compared with binary cross entropy (BCE), ASL applies different focusing strengths to active and inactive labels and suppresses easy inactive labels through probability shifting.

For one received sample, let $p_c=\operatorname{sigmoid}(o_c)$ be the predicted probability of the $c$-th interference category and $y_c\in\{0,1\}$ be the corresponding ground truth label.
The shifted negative probability is defined as
\begin{equation}
\bar{p}_c=\min(1,1-p_c+\delta),
\end{equation}
where $\delta$ is the probability shift.
The probability used in the focusing term and the focusing parameter are defined as
\begin{equation}
p^{\star}_c=y_cp_c+(1-y_c)\bar{p}_c,
\end{equation}
\begin{equation}
\gamma_c=y_c\gamma_{+}+(1-y_c)\gamma_{-},
\end{equation}
where $\gamma_{+}$ and $\gamma_{-}$ control the positive and negative focusing strengths, respectively.
Let $\langle u\rangle_{\epsilon}=\max(u,\epsilon)$, where $\epsilon$ is a small constant used for numerical stability.
The ASL for one received sample is given by
\begin{align}
\mathcal{L}_{\mathrm{ASL}}
&=-\frac{1}{C}\sum_{c=1}^{C}
\left(1-p^{\star}_c\right)^{\gamma_c}
\Big[
y_c\log\langle p_c\rangle_{\epsilon} +(1-y_c)\log\langle \bar{p}_c\rangle_{\epsilon}
\Big].
\label{eq:asl_loss}
\end{align}

\subsubsection{Category-Specific Threshold Selection}

The bag-level probabilities are converted to binary labels using category-specific thresholds.
Let $\mathcal{D}_{\mathrm{thr}}=\{(\bm{X}_i,\bm{y}_i)\}_{i=1}^{N_{\mathrm{thr}}}$ denote the threshold selection set, $p_{i,c}$ be the predicted probability of the $c$-th category for $\bm{X}_i$, and $\mathcal{T}$ be the candidate threshold set.
For a candidate threshold $\tau\in\mathcal{T}$, define $\hat{y}_{i,c}(\tau)=\mathbb{I}\left(p_{i,c}>\tau\right)$.
For each category, the threshold is selected by maximizing its F1 score as
\begin{equation}
\begin{aligned}
\tau_c^{\star}
&=\operatorname*{arg\,max}_{\tau\in\mathcal{T}}\mathrm{F1}_c(\tau)\\
&=\operatorname*{arg\,max}_{\tau\in\mathcal{T}}
\frac{2\sum_{i=1}^{N_{\mathrm{thr}}}y_{i,c}\hat{y}_{i,c}(\tau)}
{\sum_{i=1}^{N_{\mathrm{thr}}}y_{i,c}+\sum_{i=1}^{N_{\mathrm{thr}}}\hat{y}_{i,c}(\tau)},
\quad c=1,2,\ldots,C.
\end{aligned}
\label{eq:dynamic_threshold}
\end{equation}
During inference, the selected thresholds are fixed, and the predicted label is obtained as
\begin{equation}
\hat{y}_c=\mathbb{I}\left(p_c>\tau_c^{\star}\right).
\end{equation}

\subsection{Two-Stage Training Algorithm}
\label{subsec:training_algorithm}
The proposed method is trained using a two-stage procedure. 
The first stage targets single-to-compound generalization, where the model is trained with non-interference and single interference samples.
Let $\mathcal{D}_{\mathrm{train}}^{\mathrm{sing}}$ and $\mathcal{D}_{\mathrm{val}}^{\mathrm{sing}}$ denote the corresponding training and validation sets, respectively.
The trainable parameters are optimized by minimizing the ASL loss in \eqref{eq:asl_loss}, and the category-specific thresholds are selected on $\mathcal{D}_{\mathrm{val}}^{\mathrm{sing}}$ according to \eqref{eq:dynamic_threshold}.
This stage produces the single interference checkpoint $\bm{\theta}_{\mathrm{sing}}$.

The second stage targets few-shot compound interference adaptation when a limited number of labeled compound interference samples are available.
Let $\mathcal{D}_{\mathrm{train}}^{\mathrm{comp},(r)}$ denote the limited compound interference subset sampled with $r$ examples for each selected label vector and INR condition, and let $\mathcal{D}_{\mathrm{train}}^{\mathrm{sing},(r)}$ denote the corresponding subset sampled from non-interference and single interference samples.
The mixed few-shot training set is constructed as $\mathcal{D}_{\mathrm{ft,train}}^{(r)}=\mathcal{D}_{\mathrm{train}}^{\mathrm{comp},(r)}\cup\mathcal{D}_{\mathrm{train}}^{\mathrm{sing},(r)}$.
Similarly, the threshold selection set is constructed from the validation split as $\mathcal{D}_{\mathrm{ft,val}}^{(r)}=\mathcal{D}_{\mathrm{val}}^{\mathrm{comp},(r)}\cup\mathcal{D}_{\mathrm{val}}^{\mathrm{sing},(r)}$.
Starting from $\bm{\theta}_{\mathrm{sing}}$, the model is fine-tuned on $\mathcal{D}_{\mathrm{ft,train}}^{(r)}$ to adapt to compound interference while maintaining performance on non-interference and single interference samples.
Fine-tuning uses a smaller learning rate $\eta_{\mathrm{ft}}<\eta$, and the thresholds are reselected on $\mathcal{D}_{\mathrm{ft,val}}^{(r)}$ after each fine-tuning epoch.
The complete training procedure is summarized in Algorithm~\ref{alg:training}.

\begin{algorithm}[!h]
\caption{Two-Stage Training Algorithm}
\label{alg:training}
\renewcommand{\algorithmicrequire}{\textbf{Input:}}
\renewcommand{\algorithmicensure}{\textbf{Output:}}
\begin{algorithmic}[1]
\REQUIRE $\mathcal{D}_{\mathrm{train}}^{\mathrm{sing}}$; $\mathcal{D}_{\mathrm{val}}^{\mathrm{sing}}$; optional $\mathcal{D}_{\mathrm{ft,train}}^{(r)}$ and $\mathcal{D}_{\mathrm{ft,val}}^{(r)}$; $\eta$, $\eta_{\mathrm{ft}}$; $E_{\mathrm{sing}}$, $E_{\mathrm{ft}}$.
\ENSURE Model parameters $\bm{\theta}_{\mathrm{sing}}$ or $\bm{\theta}_{\mathrm{ft}}$; threshold vector $\bm{\tau}^{\star}$.
\STATE Initialize the model and freeze the pretrained DINOv3 ViT backbone;
\FOR{$e=1,\ldots,E_{\mathrm{sing}}$}
    \STATE Update $\bm{\theta}$ on $\mathcal{D}_{\mathrm{train}}^{\mathrm{sing}}$ by minimizing \eqref{eq:asl_loss} with learning rate $\eta$;
    \STATE Select $\bm{\tau}^{\star}$ on $\mathcal{D}_{\mathrm{val}}^{\mathrm{sing}}$ according to \eqref{eq:dynamic_threshold};
    \STATE Save $\bm{\theta}_{\mathrm{sing}}=\bm{\theta}$ and $\bm{\tau}^{\star}$ if the validation loss decreases;
\ENDFOR
\IF{$\mathcal{D}_{\mathrm{ft,train}}^{(r)}$ is available}
    \STATE Initialize $\bm{\theta}$ from $\bm{\theta}_{\mathrm{sing}}$;
    \FOR{$e=1,\ldots,E_{\mathrm{ft}}$}
        \STATE Update $\bm{\theta}$ on $\mathcal{D}_{\mathrm{ft,train}}^{(r)}$ by minimizing \eqref{eq:asl_loss} with learning rate $\eta_{\mathrm{ft}}$;
        \STATE Select $\bm{\tau}^{\star}$ on $\mathcal{D}_{\mathrm{ft,val}}^{(r)}$ according to \eqref{eq:dynamic_threshold};
        \STATE Save $\bm{\theta}_{\mathrm{ft}}=\bm{\theta}$ and $\bm{\tau}^{\star}$ if the validation loss decreases;
    \ENDFOR
\ENDIF
\end{algorithmic}
\end{algorithm}
\vspace{-1em}

\section{Experiments and Results}
\label{sec:experiments}

\subsection{Experimental Setup}
\label{subsec:experimental_setup}
\subsubsection{Dataset Construction}
The experimental dataset is generated at complex baseband by using LR-FHSS uplink waveforms as background communication signals and adding interference components.
The LR-FHSS waveforms follow the frame structure and hopping mechanism in \cite{lrfhss-overview}.
The satellite uplink channel includes time-varying Doppler, shadowed-Rician slow fading, and receiver AWGN, following LR-FHSS direct-to-satellite studies \cite{dts-lrfhss-simulation,lrfhss-outage}.

For each received sample, the multi-hot label vector records the active interference categories, and the all zero vector represents a non-interference sample.
Let $I$ denote the number of active interference components in a sample, where $I\in\{0,1,2,3\}$.
The INR follows the definition in Section~\ref{sec:system_model}, i.e., $\mathrm{INR}=10\log_{10}(P_{\mathrm{ref}}/\sigma_w^2)$, where $P_{\mathrm{ref}}$ is the reference power of a single interference component.
Each active interference component is scaled to $P_{\mathrm{ref}}$, and the INR is used to set the noise variance $\sigma_w^2$.
For $I=0$, no interference component is added, but the same reference power $P_{\mathrm{ref}}$ is used to set $\sigma_w^2$.

The time-frequency image and frequency sequence are generated from each received signal as follows.
For the time-frequency image, the STFT power spectrogram is reduced to a $224\times224$ matrix by max pooling over the frequency and time axes.
The result is converted to the dB scale and min--max normalized to the 8-bit range for each image.
The Plasma colormap is applied to obtain an RGB image.
The image is vertically flipped so that lower frequencies appear at the bottom, yielding
$\bm{X}^{\mathrm{tf}}\in\mathbb{R}^{3\times224\times224}$.
For the frequency sequence, the original STFT power spectrogram is max pooled over the time axis and converted to relative dB values with its peak normalized to $0$~dB, yielding
$\bm{X}^{\mathrm{f}}\in\mathbb{R}^{8192}$.

\begin{figure}[!t]
\centering
\begin{minipage}[t]{0.45\columnwidth}
\centering
\includegraphics[width=\linewidth]{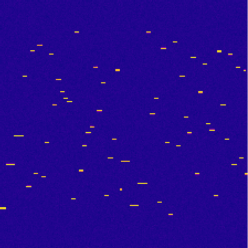}\\[0.2ex]
\footnotesize (a) $I=0$ (LR-FHSS signal only)\\[0.15ex]
\footnotesize $\bm{y}=0000000$
\end{minipage}
\hfill
\begin{minipage}[t]{0.45\columnwidth}
\centering
\includegraphics[width=\linewidth]{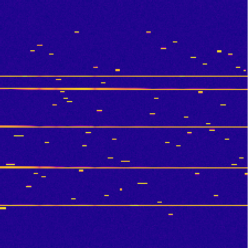}\\[0.2ex]
\footnotesize (b) $I=1$ (TI)\\[0.15ex]
\footnotesize $\bm{y}=1000000$
\end{minipage}
\vspace{0.7ex}
\begin{minipage}[t]{0.45\columnwidth}
\centering
\includegraphics[width=\linewidth]{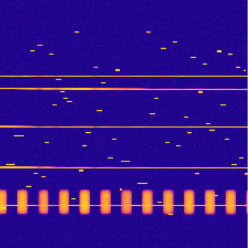}\\[0.2ex]
\footnotesize (c) $I=2$ (TI+PPNI)\\[0.15ex]
\footnotesize $\bm{y}=1010000$
\end{minipage}
\hfill
\begin{minipage}[t]{0.45\columnwidth}
\centering
\includegraphics[width=\linewidth]{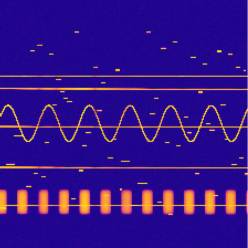}\\[0.2ex]
\footnotesize (d) $I=3$ (TI+PPNI+SFMI)\\[0.15ex]
\footnotesize $\bm{y}=1010100$
\end{minipage}
\caption{Representative time-frequency images for $I=0,1,2,3$. The four examples use the same LR-FHSS background realization for visual comparison, and the label-vector order is TI, PBNI, PPNI, LFMI, SFMI, RTI, and RCHI.}
\label{fig:interference_cardinality}
\end{figure}

Fig.~\ref{fig:interference_cardinality} illustrates representative time-frequency images with different numbers of active interference components.
The main simulation parameters are summarized in Table~\ref{tab:lrfhss_params}, and the parameters of the seven interference categories defined in Section~\ref{sec:system_model} are summarized in Table~\ref{tab:interference_types}.
The generated LR-FHSS compound interference dataset and its documentation are publicly available at Zenodo\footnote{\url{https://doi.org/10.5281/zenodo.21298517}}.

\begin{table}[!t]
\caption{Main Simulation Parameters}
\label{tab:lrfhss_params}
\centering
\scriptsize
\renewcommand{\arraystretch}{1.05}
\begin{tabular}{@{}p{0.5\linewidth}@{}p{0.5\linewidth}@{}}
\toprule
Parameter & Value \\
\midrule
Region & US902--928 / US915 \\
Operating channel width (OCW) & 1.5234375 MHz \\
Selected OCW center frequency & 903.0 MHz \\
Grid spacing $\Delta f_g$ & 25390.625 Hz \\
Hop grid points per OCW & 60 \\
Data rate & DR5 \\
Coding rate & 1/3 \\
PHY header copies & 3 \\
Header hop PHY samples & 114 \\
Payload fragments & 27 \\
Payload hop PHY samples & 50 \\
PHY sample rate & 488.28125 Hz \\
Sampling rate $f_s$ & 2 MHz \\
Samples per PHY sample & 4096 \\
Frame duration & 3.403776 s \\
Observation duration & 6.807552 s \\
STFT window & Blackman \\
FFT size $N_{\mathrm{fft}}$ / hop size $R$ & 8192 / 8192 \\
Image size $H\times W$ & $224\times224$ \\
Doppler offset per source & $[-20,20]$ kHz \\
Doppler rate per source & $[-500,500]$ Hz/s \\
Shadowed-Rician update & 0.5 s \\
Shadowed-Rician $K$ / $m$ & $[4,12]$ dB / $[1,5]$ \\
Reference interference power $P_{\mathrm{ref}}$ & 1.0 \\
ISR & 0 dB \\
INR range & $[-30,10]$ dB, step 2 dB \\
Samples per label vector and INR & 100 \\
Dataset split & Train/val/test $=0.6/0.2/0.2$ \\
Fine-tuning sample count $r$ & $\{0,1,2,3\}$ \\
\bottomrule
\end{tabular}
\vspace{-0em}
\end{table}

\begin{table}[!t]
\caption{Interference Parameters}
\label{tab:interference_types}
\centering
\scriptsize

\begin{tabular}{@{}m{0.30\linewidth}@{}>{\raggedright\arraybackslash}m{0.70\linewidth}@{}}
\toprule
Interference & Parameters \\
\midrule
\multirow{3}{*}{TI} & $U\in\{1,\ldots,5\}$\\
& $f_u\in\{-30,\ldots,29\}\Delta f_g$\\
& $\phi_u\in[0,2\pi)$ \\
\midrule
PBNI & $B\in[76.171875,457.03125]$ kHz\\
\midrule
\multirow{3}{*}{PPNI} & $B\in[76.171875,304.6875]$ kHz\\
& $T_p\in[0.49152,1.31072]$ s\\
& $\tau_p\in\{0.3,0.5,0.7\}$ \\
\midrule
\multirow{2}{*}{LFMI} & $B_{\mathrm{LFMI}}\in[152.34375,761.71875]$ kHz\\
& $N_{\mathrm{LFMI}}/f_s\in[0.49152,1.31072]$ s \\
\midrule
\multirow{2}{*}{SFMI} & $\Delta f_{\mathrm{SFMI}}\in[38.0859375,228.515625]$ kHz\\
& $N_{\mathrm{SFMI}}/f_s\in[0.49152,1.31072]$ s \\
\midrule
RTI & $d\in[0.053248,0.081920]$ s\\
\midrule
\multirow{2}{*}{RCHI} & $d\in[0.053248,0.081920]$ s\\
& $N_{\mathrm{cap}}/f_s\in[0.3072,0.7168]$ s \\
\bottomrule
\end{tabular}
\vspace{-0em}
\end{table}

\subsubsection{Implementation Details}
\label{subsubsec:implementation_details}
Two experimental scenarios are considered according to the availability of labeled compound interference samples.

In the first scenario, models are trained on $\mathcal{D}_{\mathrm{train}}^{\mathrm{sing}}$ and validated on $\mathcal{D}_{\mathrm{val}}^{\mathrm{sing}}$, where $I=0,1$.
This scenario corresponds to single-to-compound generalization, in which only non-interference and single interference samples are available during training, whereas compound interference samples may appear during operation.
After each training epoch, the threshold vector $\bm{\tau}^{\star}$ is selected on $\mathcal{D}_{\mathrm{val}}^{\mathrm{sing}}$ according to \eqref{eq:dynamic_threshold}, with $\mathcal{T}=\{0,0.01,\ldots,1\}$.
The trained models are evaluated separately on test subsets with $I=0,1,2,3$.

In the second scenario, the model trained on non-interference and single interference samples is further fine-tuned on $\mathcal{D}_{\mathrm{ft,train}}^{(r)}$, which combines $\mathcal{D}_{\mathrm{train}}^{\mathrm{comp},(r)}$ with $\mathcal{D}_{\mathrm{train}}^{\mathrm{sing},(r)}$.
This scenario corresponds to few-shot compound interference adaptation, where a small number of labeled compound interference samples become available through receiver monitoring or offline analysis.
Here, $r\in\{0,1,2,3\}$ is the number of samples selected for each label vector and INR condition from both the training and validation splits, corresponding to 0, 1344, 2688, and 4032 samples in each of $\mathcal{D}_{\mathrm{ft,train}}^{(r)}$ and $\mathcal{D}_{\mathrm{ft,val}}^{(r)}$, respectively.
After each fine-tuning epoch, $\bm{\tau}^{\star}$ is reselected on $\mathcal{D}_{\mathrm{ft,val}}^{(r)}$ using the same threshold set $\mathcal{T}$.
The fine-tuned models are evaluated on the combined test set with $I=0,1,2,3$.

All models use the same implementation hyperparameters unless otherwise specified.
For all training and fine-tuning runs, the ASL hyperparameters in \eqref{eq:asl_loss} are fixed as $\delta=0.05$, $\gamma_{+}=0$, $\gamma_{-}=4$, and $\epsilon=10^{-8}$.
The batch size is 32, and the optimizer is Adam with weight decay $10^{-4}$.
In both scenarios, the learning rate is reduced by a factor of 0.1 using a ReduceLROnPlateau scheduler with patience 3.
Single interference training uses a learning rate of $10^{-3}$ for at most 100 epochs with early stopping patience 6, whereas few-shot fine-tuning uses a lower learning rate of $10^{-4}$ for 10 epochs.
To ensure statistical reliability, each reported result is averaged over five independent runs with different random seeds, and the same random seeds are used for all compared models under each experimental scenario.

For the proposed method, Table~\ref{tab:aux_frequency_branch} summarizes the network architecture of the frequency branch used to implement $f_{\mathrm{aux}}(\cdot)$.
The channel attention operation follows the efficient channel attention mechanism in \cite{eca-net} and is adapted to one-dimensional frequency features.
The MBConv1d blocks follow a one-dimensional adaptation of the MobileNetV2 inverted residual design \cite{mobilenetv2}.
The projection layer in each branch is implemented as a single linear layer, and the common output dimension is set to $D_{\mathrm{o}}=768$.
The shared instance classifier $f_{\mathrm{cls}}(\cdot)$ comprises three linear layers with hidden dimension 384, together with layer normalization, GELU activation, and dropout.
Only the projection layers, the frequency branch, and the shared instance classifier are optimized during training, while the pretrained DINOv3 ViT backbone is kept frozen.
This avoids retraining the large pretrained backbone.

\begin{table}[!h]
\caption{Network Architecture of the Frequency Branch}
\label{tab:aux_frequency_branch}
\centering
\scriptsize
\setlength{\tabcolsep}{3pt}
\begin{tabular}{@{}>{\raggedright\arraybackslash}p{0.27\columnwidth}>{\raggedright\arraybackslash}p{0.43\columnwidth}>{\centering\arraybackslash}p{0.22\columnwidth}@{}}
\toprule
Operation & Configuration & Output size \\
\midrule
Input & Max-compressed frequency sequence & $F$ \\
Conv1d-BN-ReLU & $1\rightarrow64$, $k=7$, $s=2$ & $64\times F/2$ \\
Channel attention & $64\rightarrow64$, $k=3$ & $64\times F/2$ \\
MBConv1d-1 & $64\rightarrow128$, $k=5$, $s=2$ & $128\times F/4$ \\
MBConv1d-2 & $128\rightarrow256$, $k=5$, $s=2$ & $256\times F/8$ \\
MBConv1d-3 & $256\rightarrow512$, $k=5$, $s=2$ & $512\times F/16$ \\
AdaptiveAvgPool1d & $F/16\rightarrow T_{\mathrm{f}}=8$ & $T_{\mathrm{f}}\times512$ \\
Conv1d-BN-GELU-1 & $512\rightarrow512$, $k=3$ & $T_{\mathrm{f}}\times512$ \\
Conv1d-BN-GELU-2 & $512\rightarrow512$, $k=3$ & $T_{\mathrm{f}}\times512$ \\
\bottomrule
\end{tabular}
\vspace{-0em}
\end{table}

To match the input format and distribution of the pretrained DINOv3 backbone, each RGB time-frequency image is scaled from the 8-bit range to $[0,1]$.
The RGB channels are then normalized using the ImageNet mean $[0.485,0.456,0.406]$ and standard deviation $[0.229,0.224,0.225]$.

\subsubsection{Evaluation Metrics}
We report exact accuracy and label accuracy for multi-label compound interference recognition.
Let $\mathcal{D}_{\mathrm{test}}^{(I)}=\{(\bm{X}_i,\bm{y}_i)\}_{i=1}^{N_I}$ denote the test subset with $I$ active interference components, and let $\mathcal{I}$ denote the set of values of $I$ included in the overall evaluation.
The total number of evaluated samples is $N_{\mathrm{test}}=\sum_{I\in\mathcal{I}}N_I$.
Let $\hat{\bm{y}}_i$ be the thresholded multi-hot prediction.

The exact accuracy on $\mathcal{D}_{\mathrm{test}}^{(I)}$ is defined as
\begin{equation}
\mathrm{Acc}_{\mathrm{exact}}^{(I)}
=\frac{1}{N_I}\sum_{i=1}^{N_I}\mathbb{I}\left(\hat{\bm{y}}_i=\bm{y}_i\right).
\label{eq:exact_accuracy}
\end{equation}
For the combined test set, the overall exact accuracy is computed as
\begin{equation}
\mathrm{Acc}_{\mathrm{exact}}
=\frac{1}{N_{\mathrm{test}}}\sum_{I\in\mathcal{I}}N_I\mathrm{Acc}_{\mathrm{exact}}^{(I)}.
\label{eq:overall_exact_accuracy}
\end{equation}

The label accuracy on $\mathcal{D}_{\mathrm{test}}^{(I)}$ is defined as
\begin{equation}
\mathrm{Acc}_{\mathrm{label}}^{(I)}
=\frac{1}{N_I C}\sum_{i=1}^{N_I}\sum_{c=1}^{C}
\mathbb{I}\left(\hat{y}_{i,c}=y_{i,c}\right).
\label{eq:label_accuracy}
\end{equation}
Similarly, the overall label accuracy is computed as
\begin{equation}
\mathrm{Acc}_{\mathrm{label}}
=\frac{1}{N_{\mathrm{test}}}\sum_{I\in\mathcal{I}}N_I\mathrm{Acc}_{\mathrm{label}}^{(I)}.
\label{eq:overall_label_accuracy}
\end{equation}

The exact accuracy is a more stringent metric because it requires all entries of the multi-hot label vector to be correctly predicted, whereas the label accuracy measures the average correctness of binary decisions for each interference category.

For category-wise recognition difficulty analysis, the recall of interference category $c$ on $\mathcal{D}_{\mathrm{test}}^{(I)}$ is defined for test subsets containing positive samples of category $c$ as
\begin{equation}
\mathrm{Recall}_{c}^{(I)}
=
\frac{
\sum_{i=1}^{N_I}\mathbb{I}\left(y_{i,c}=1\right)\mathbb{I}\left(\hat{y}_{i,c}=1\right)
}{
\sum_{i=1}^{N_I}\mathbb{I}\left(y_{i,c}=1\right)
}.
\label{eq:category_wise_recall}
\end{equation}

\begin{table*}[!b]
\caption{Recognition Performance under Single-to-Compound Generalization with $I=0,1,2,3$}
\label{tab:main_results}
\centering
\scriptsize
\setlength{\tabcolsep}{2.6pt}
\resizebox{\textwidth}{!}{%
\begin{tabular}{>{\raggedright\arraybackslash}m{0.14\textwidth}cccccccccc}
\toprule
Model & $\mathrm{Acc}_{\mathrm{exact}}^{(0)}$ (\%) & $\mathrm{Acc}_{\mathrm{label}}^{(0)}$ (\%) & $\mathrm{Acc}_{\mathrm{exact}}^{(1)}$ (\%) & $\mathrm{Acc}_{\mathrm{label}}^{(1)}$ (\%) & $\mathrm{Acc}_{\mathrm{exact}}^{(2)}$ (\%) & $\mathrm{Acc}_{\mathrm{label}}^{(2)}$ (\%) & $\mathrm{Acc}_{\mathrm{exact}}^{(3)}$ (\%) & $\mathrm{Acc}_{\mathrm{label}}^{(3)}$ (\%) & $\mathrm{Acc}_{\mathrm{exact}}$ (\%) & $\mathrm{Acc}_{\mathrm{label}}$ (\%) \\
\midrule
MLAMC & 94.05 & 99.15 & 92.38 & 98.78 & 21.44 & 86.42 & 0.66 & 71.59 & 18.97 & 79.86 \\
MIML-VGG16 & 93.81 & 99.12 & 90.99 & 98.60 & 28.14 & 88.62 & 1.83 & 76.71 & 21.65 & 83.36 \\
MIRNet & 97.62 & 99.66 & \textbf{95.78} & \textbf{99.32} & 16.52 & 85.25 & 0.12 & 70.30 & 17.49 & 78.84 \\
Proposed (ViT-S/16) & 94.05 & 99.15 & 94.90 & 99.12 & \textbf{51.94} & \textbf{92.77} & 13.58 & \textbf{84.04} & 36.32 & \textbf{88.79} \\
Proposed (ViT-S+/16) & \textbf{98.33} & \textbf{99.76} & 94.86 & 99.17 & 42.32 & 91.32 & 8.90 & 82.15 & 30.67 & 87.30 \\
Proposed (ViT-B/16) & 97.38 & 99.63 & 95.68 & 99.28 & 49.40 & 92.42 & \textbf{14.93} & 84.03 & \textbf{36.36} & 88.69 \\
\bottomrule
\end{tabular}%
}
\vspace{-1em}
\end{table*}

\begin{figure*}[!t]
\centering
\begin{minipage}[t]{0.49\textwidth}
\centering
\subfloat[$I=0$]{%
\includegraphics[width=0.50\linewidth]{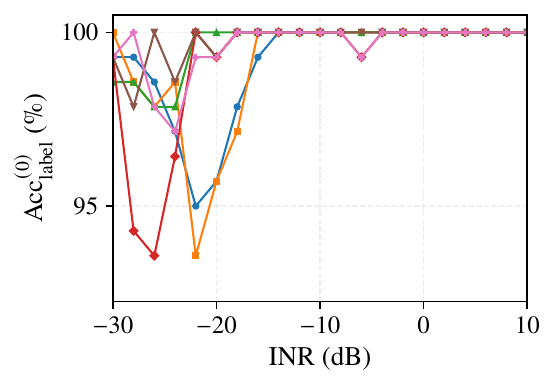}}
\hfill
\subfloat[$I=1$]{%
\includegraphics[width=0.50\linewidth]{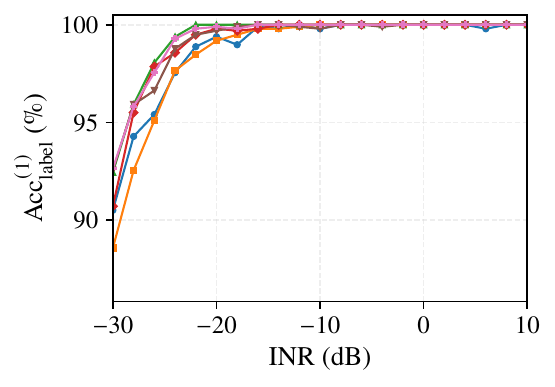}}
\\[-0.4ex]
\subfloat[$I=2$]{%
\includegraphics[width=0.50\linewidth]{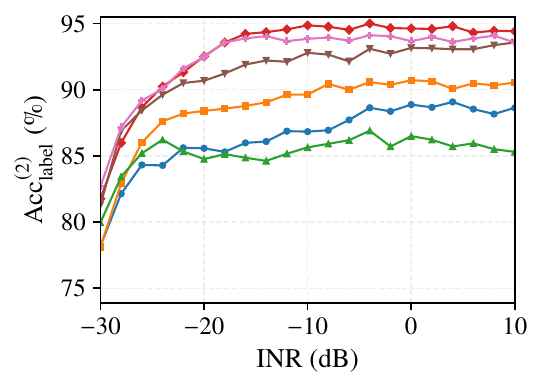}}
\hfill
\subfloat[$I=3$]{%
\includegraphics[width=0.50\linewidth]{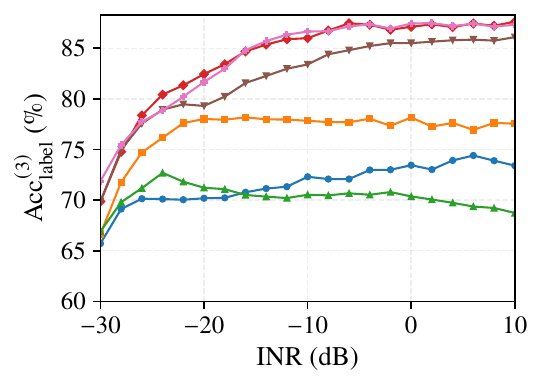}}
\\[-0.4ex]
\includegraphics[width=0.95\linewidth]{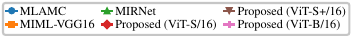}
\caption{Label accuracy versus INR under single-to-compound generalization with $I=0,1,2,3$.}
\label{fig:label_acc_inr}
\end{minipage}
\hfill
\begin{minipage}[t]{0.49\textwidth}
\centering
\subfloat[$I=0$]{%
\includegraphics[width=0.50\linewidth]{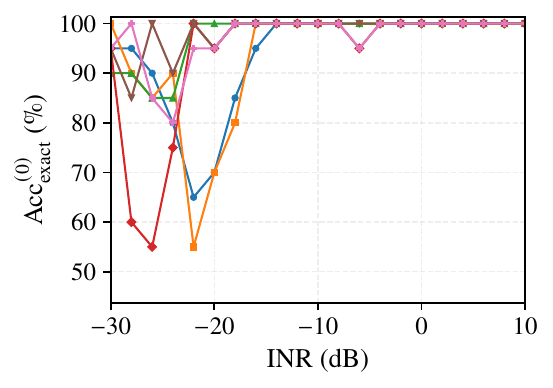}}
\hfill
\subfloat[$I=1$]{%
\includegraphics[width=0.50\linewidth]{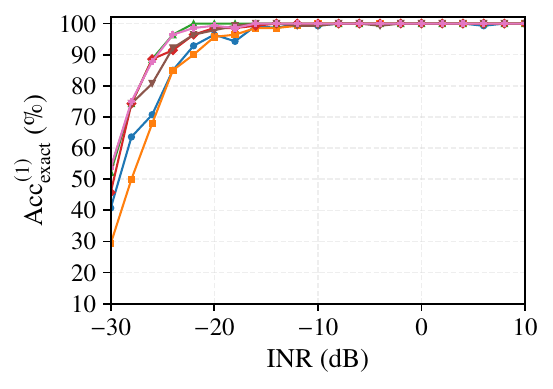}}
\\[-0.4ex]
\subfloat[$I=2$]{%
\includegraphics[width=0.50\linewidth]{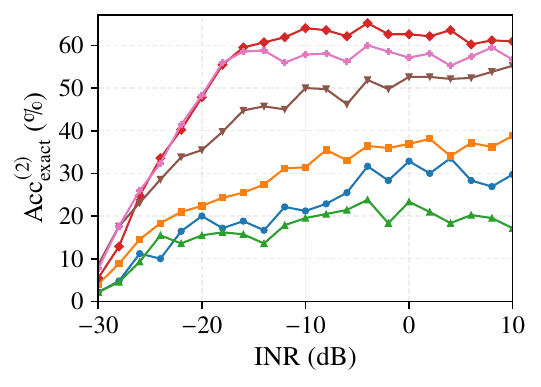}}
\hfill
\subfloat[$I=3$]{%
\includegraphics[width=0.50\linewidth]{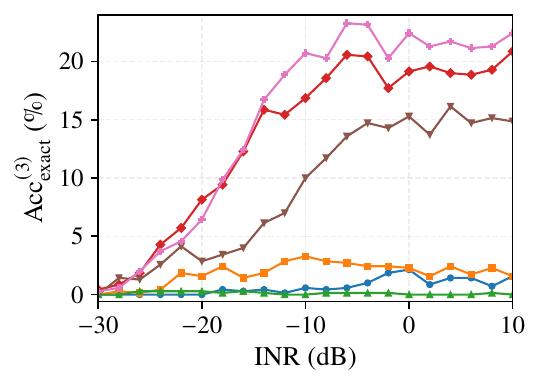}}
\\[-0.4ex]
\includegraphics[width=0.95\linewidth]{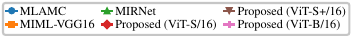}
\caption{Exact accuracy versus INR under single-to-compound generalization with $I=0,1,2,3$.}
\label{fig:exact_acc_inr}
\end{minipage}
\vspace{-0em}
\end{figure*}

\subsubsection{Compared Recognition Models}
The comparison includes three representative baseline models for compound interference recognition.
MLAMC \cite{compound-signal-mlamc} is used as a CNN-based multi-label learning baseline.
MIML-VGG16 \cite{overlapping-lpi-miml} is used as a multi-instance multi-label learning baseline with VGG16 as the backbone.
MIRNet \cite{compound-interference-uav} is used as a multimodal multi-label learning baseline that combines time-frequency images and frequency sequences.
For the proposed method, we evaluate three DINOv3 ViT variants with increasing backbone scale, namely ViT-S/16, ViT-S+/16, and ViT-B/16 \cite{vit,dinov3}.
Unless otherwise specified, all baseline and proposed models are trained and evaluated using the same data splits, evaluation metrics, threshold selection strategy, and ASL loss function.

\subsection{Single-to-Compound Generalization}
\label{subsec:single_to_compound_generalization}
Following the first scenario in Section~\ref{subsec:experimental_setup}, this subsection evaluates single-to-compound generalization, where models trained only on non-interference and single interference samples are directly tested on compound interference samples without fine-tuning.

\subsubsection{Overall Recognition Performance}
Table~\ref{tab:main_results} reports the exact and label accuracies on the $I=0,1,2,3$ test subsets and on the combined test set.
The best result for each metric is highlighted in bold.
The proposed ViT-B/16 and ViT-S/16 variants achieve the highest overall exact and label accuracies of 36.36\% and 88.79\%, respectively, outperforming the strongest baseline MIML-VGG16 by 14.71 and 5.43 percentage points.
For $I=0$ and $I=1$, all compared models achieve high exact and label accuracies, indicating reliable recognition under the interference conditions included in training.
MIRNet obtains the best results for $I=1$, whereas the proposed ViT-S+/16 variant obtains the best results for $I=0$.

\begin{figure*}[!t]
\centering
\includegraphics[width=0.90\textwidth]{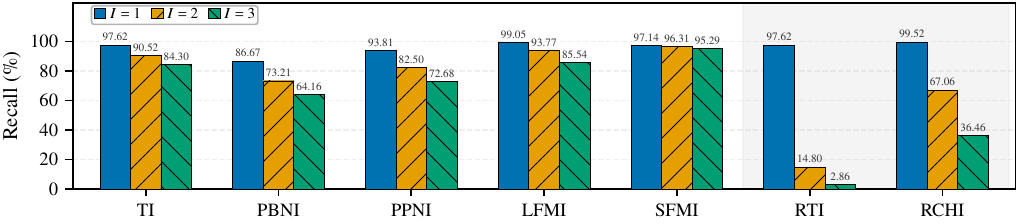}
\caption{Category-wise recall of the proposed ViT-B/16 model on the $I=1$, $I=2$, and $I=3$ test subsets under single-to-compound generalization. The shaded categories denote reactive interference categories.}
\label{fig:category_wise_recall}
\vspace{-0em}
\end{figure*} 

The advantage of the proposed method becomes more evident on the unseen compound interference subsets.
For $I=2$, the proposed ViT-S/16 variant achieves the highest exact and label accuracies of 51.94\% and 92.77\%, respectively, outperforming MIML-VGG16 by 23.80 and 4.15 percentage points.
For $I=3$, the proposed ViT-B/16 and ViT-S/16 variants achieve the highest exact and label accuracies of 14.93\% and 84.04\%, respectively, outperforming MIML-VGG16 by 13.10 and 7.33 percentage points.
These results indicate that the proposed method can better exploit localized interference evidence learned from non-interference and single interference samples.
By aggregating local instances from the time-frequency and frequency domains, the model can identify active interference components even when their combinations are unseen during training.
Nevertheless, exact recognition remains challenging when multiple unseen interference components coexist.

Figs.~\ref{fig:label_acc_inr} and \ref{fig:exact_acc_inr} show the label and exact accuracies versus INR, respectively.
For $I=0$ and $I=1$, both metrics increase rapidly with INR and remain high when the INR is at least $-16$~dB.
For the unseen compound subsets with $I=2$ and $I=3$, the proposed models generally outperform the baselines across INR values, with a more pronounced advantage when the INR is at least $-20$~dB.
As $I$ increases, the exact accuracy decreases more noticeably than the label accuracy, reflecting the stricter criterion of exact accuracy.
This indicates that models may still identify some active components correctly, but complete recovery of all coexisting interference components becomes increasingly difficult.

\subsubsection{Category-Wise Recall Analysis}
Fig.~\ref{fig:category_wise_recall} reports the category-wise recall of the proposed ViT-B/16 model on the $I=1,2,3$ test subsets.
The recall is computed according to \eqref{eq:category_wise_recall}.
Since all label vectors are uniformly represented in each test subset, each interference category appears 420, 2520, and 6300 times for $I=1$, $I=2$, and $I=3$, respectively.

For $I=1$, all interference categories achieve high recall, confirming that each interference category can be reliably identified when it appears alone.
As $I$ increases to 2 and 3, the recall decreases unevenly across categories.
The RTI recall drops from 97.62\% for $I=1$ to 14.80\% and 2.86\% for $I=2$ and $I=3$, respectively, while the RCHI recall decreases to 67.06\% and 36.46\%.
This indicates that reactive interference categories are a major source of single-to-compound generalization degradation.
Because RTI and RCHI are coupled with the LR-FHSS hopping behavior, their localized interference evidence may be confused with LR-FHSS hopping fragments or weakened by coexisting interference components.
Therefore, labeled compound interference samples are needed to improve reactive interference recognition, motivating the few-shot compound interference adaptation scenario analyzed next.

\begin{figure*}[!h]
\centering
\subfloat[Exact accuracy]{%
\includegraphics[width=0.95\textwidth]{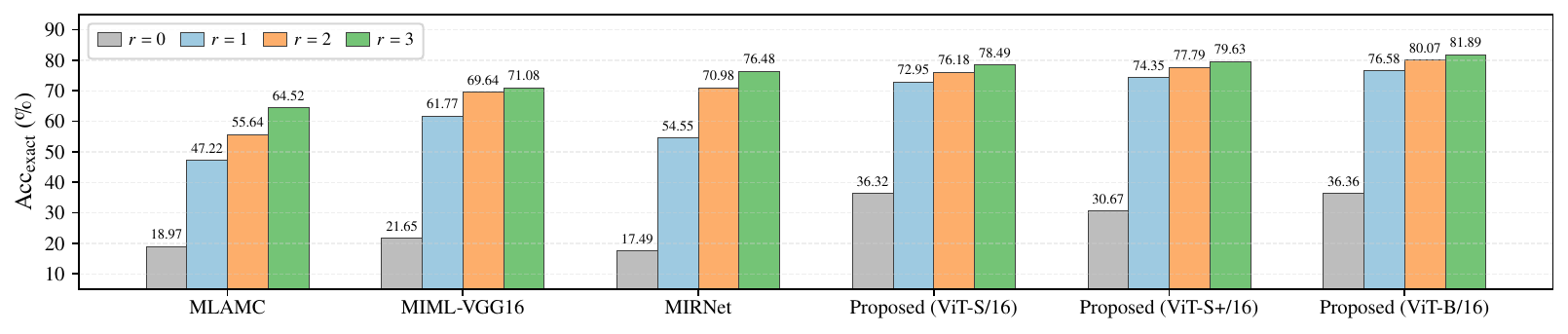}}
\\[-0.4ex]
\subfloat[Label accuracy]{%
\includegraphics[width=0.95\textwidth]{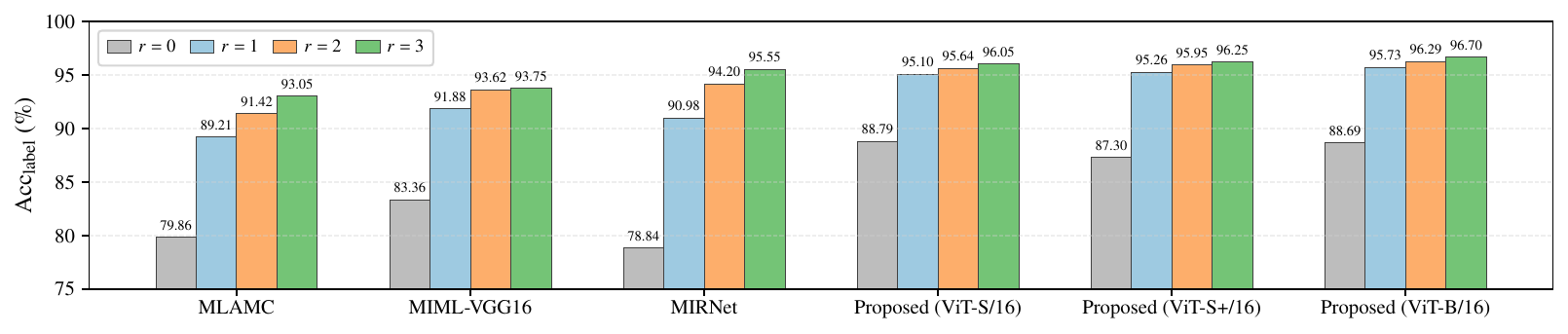}}
\caption{Overall exact and label accuracies for $r=0,1,2,3$ under few-shot compound interference adaptation.}
\label{fig:fine_tuning_overall}
\vspace{-1em}
\end{figure*}

\begin{figure*}[!t]
\centering
\begin{minipage}[t]{0.49\textwidth}
\centering
\subfloat[$r=0$]{%
\includegraphics[width=0.50\linewidth]{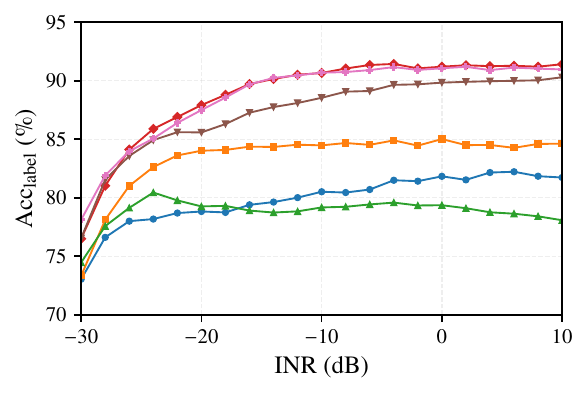}}
\hfill
\subfloat[$r=1$]{%
\includegraphics[width=0.50\linewidth]{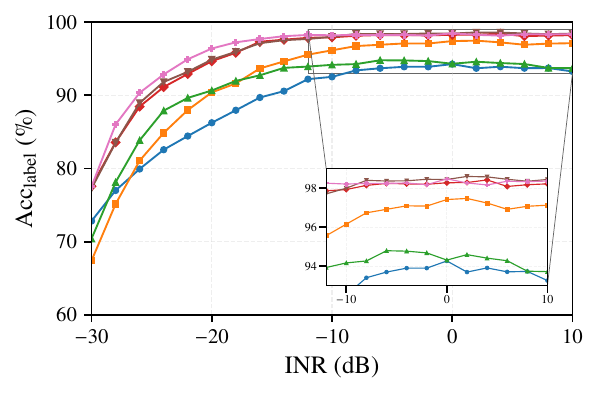}}
\\[-0.4ex]
\subfloat[$r=2$]{%
\includegraphics[width=0.50\linewidth]{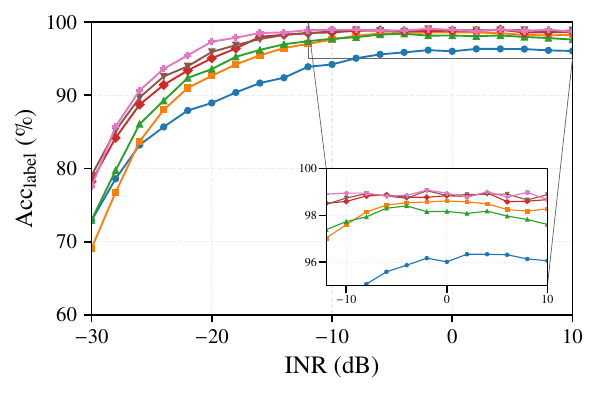}}
\hfill
\subfloat[$r=3$]{%
\includegraphics[width=0.50\linewidth]{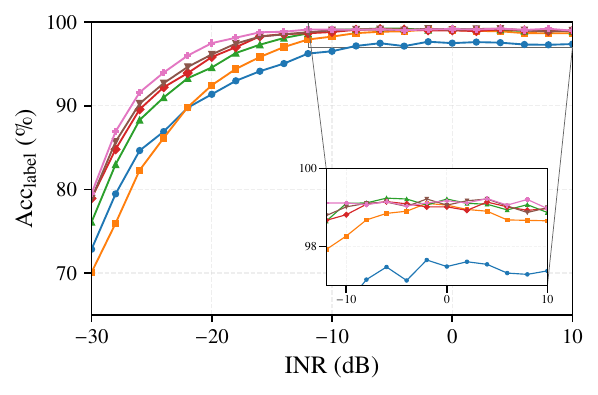}}
\\[-0.4ex]
\includegraphics[width=0.95\linewidth]{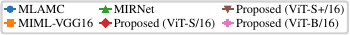}
\caption{Label accuracy versus INR for $r=0,1,2,3$ under few-shot compound interference adaptation.}
\label{fig:fine_tuning_label_acc_inr}
\end{minipage}
\hfill
\begin{minipage}[t]{0.49\textwidth}
\centering
\subfloat[$r=0$]{%
\includegraphics[width=0.50\linewidth]{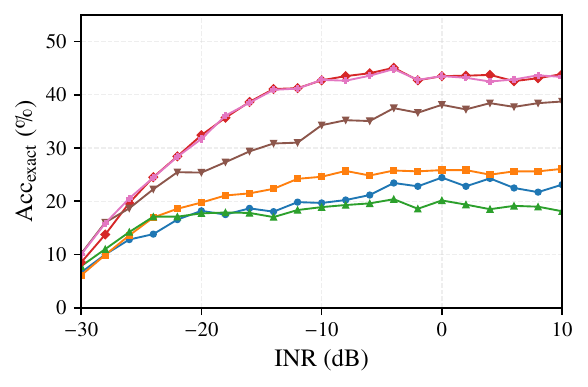}}
\hfill
\subfloat[$r=1$]{%
\includegraphics[width=0.50\linewidth]{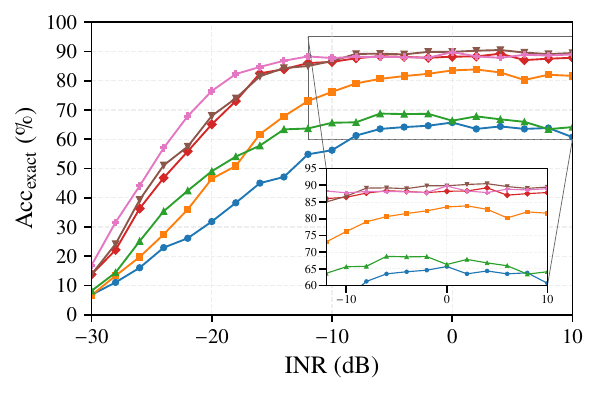}}
\\[-0.4ex]
\subfloat[$r=2$]{%
\includegraphics[width=0.50\linewidth]{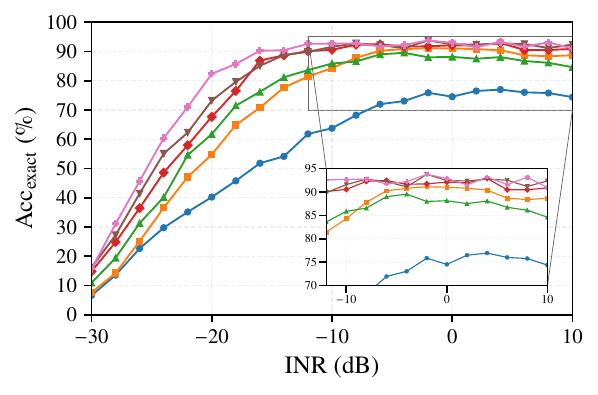}}
\hfill
\subfloat[$r=3$]{%
\includegraphics[width=0.50\linewidth]{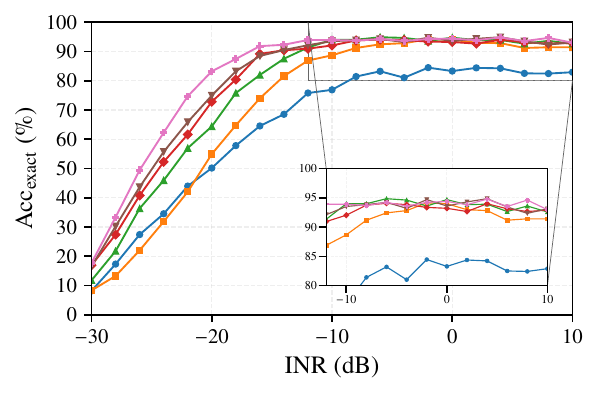}}
\\[-0.4ex]
\includegraphics[width=0.95\linewidth]{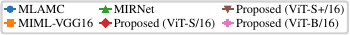}
\caption{Exact accuracy versus INR for $r=0,1,2,3$ under few-shot compound interference adaptation.}
\label{fig:fine_tuning_exact_acc_inr}
\end{minipage}
\vspace{-0em}
\end{figure*}

\begin{figure*}[!t]
\centering
\includegraphics[width=0.95\textwidth]{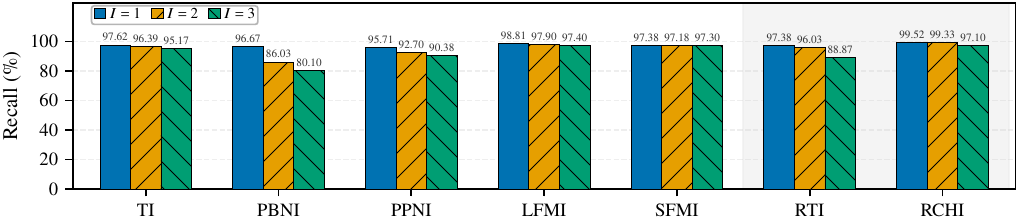}
\caption{Category-wise recall of the proposed ViT-B/16 model on the $I=1$, $I=2$, and $I=3$ test subsets under few-shot compound interference adaptation for $r=3$. The shaded categories denote reactive interference categories.}
\label{fig:few_shot_category_wise_recall}
\vspace{-0em}
\end{figure*}

\subsection{Few-Shot Compound Interference Adaptation}
Following the second scenario in Section~\ref{subsec:experimental_setup}, this subsection evaluates few-shot compound interference adaptation, where the model trained on non-interference and single interference samples is fine-tuned with a small number of labeled compound interference samples.
Here, $r=0$ denotes evaluation of $\bm{\theta}_{\mathrm{sing}}$ without fine-tuning.

\subsubsection{Overall Recognition Performance}
Fig.~\ref{fig:fine_tuning_overall} reports the overall exact and label accuracies on the combined test set with $I=0,1,2,3$ for $r=0,1,2,3$.
All compared models benefit from few-shot fine-tuning, especially in exact accuracy, showing that a small number of labeled compound interference samples helps recover complete multi-label predictions.
For $r=1$, the proposed ViT-B/16 variant achieves the best overall performance, outperforming the strongest baseline MIML-VGG16 by 14.81 and 3.85 percentage points in exact and label accuracies, respectively.
This result indicates that the proposed multi-domain instance fusion remains effective when only very limited compound interference labels are available.

As $r$ increases, MIRNet gradually surpasses MIML-VGG16 and approaches the proposed variants, suggesting that complementary frequency-domain information can be better exploited when more labeled compound interference samples are available.
The contribution of frequency-domain instances is further examined in Section~\ref{subsec:domain_contribution}.
Among the proposed variants, the overall performance improves as the frozen DINOv3 backbone scales from ViT-S/16 to ViT-B/16, indicating that stronger local representations further improve few-shot compound interference adaptation.

Figs.~\ref{fig:fine_tuning_label_acc_inr} and \ref{fig:fine_tuning_exact_acc_inr} show the label and exact accuracies versus INR for different values of $r$.
For $r=0$, the results are consistent with the single-to-compound generalization performance reported in Section~\ref{subsec:single_to_compound_generalization}.
For $r>0$, the MIRNet curves improve as $r$ increases, while the proposed variants remain above the baselines in the high INR region.
These results show that few-shot compound interference adaptation improves recognition performance, and that exploiting localized evidence from multiple signal domains remains effective after fine-tuning.

\subsubsection{Category-Wise Recall Analysis}
Fig.~\ref{fig:few_shot_category_wise_recall} reports the category-wise recall of the proposed ViT-B/16 model under few-shot compound interference adaptation with $r=3$.
For $I=1$, all interference categories remain above 95.71\%, indicating that fine-tuning preserves reliable single interference recognition.
For the compound subsets, the RTI and RCHI recalls increase to 96.03\% and 99.33\% for $I=2$, and to 88.87\% and 97.10\% for $I=3$, respectively.
Compared with Fig.~\ref{fig:category_wise_recall}, these results show that a small number of labeled compound interference samples can largely restore the recall of RTI and RCHI in compound interference samples.

\subsection{Domain Contribution Analysis}
\label{subsec:domain_contribution}
This subsection evaluates the contribution of each signal domain in the proposed multi-domain instance fusion method.
The TF variant uses only time-frequency instances extracted by the DINOv3 ViT backbone, the F variant uses only frequency instances extracted by the auxiliary network, and the TF+F variant uses instances from both domains.
All variants are trained and evaluated in the representative INR range from $-10$ to $10$~dB with a step of $4$~dB, where interference structures are sufficiently observable for comparing domain contributions.
Table~\ref{tab:domain_contribution} reports the results on the combined test set with $I=0,1,2,3$.
The columns without fine-tuning correspond to the single-to-compound generalization scenario, whereas the columns with fine-tuning correspond to few-shot compound interference adaptation for $r=3$.

\begin{table}[!h]
\caption{Domain Contribution Analysis on the Combined Test Set with $I=0,1,2,3$}
\label{tab:domain_contribution}
\centering
\scriptsize
\setlength{\tabcolsep}{4.5pt}
\begin{tabular}{@{}lcccc@{}}
\toprule
Variant & \multicolumn{2}{c}{Without fine-tuning} & \multicolumn{2}{c}{With fine-tuning} \\
\cmidrule(lr){2-3}\cmidrule(lr){4-5}
& $\mathrm{Acc}_{\mathrm{exact}}$ (\%) & $\mathrm{Acc}_{\mathrm{label}}$ (\%) & $\mathrm{Acc}_{\mathrm{exact}}$ (\%) & $\mathrm{Acc}_{\mathrm{label}}$ (\%) \\
\midrule
TF & 50.42 & 91.95 & 91.65 & 98.76 \\
F & 22.21 & 82.91 & 36.18 & 87.90 \\
TF+F & \textbf{60.18} & \textbf{93.57} & \textbf{93.20} & \textbf{99.00} \\
\bottomrule
\end{tabular}
\vspace{-0em}
\end{table}

Across both scenarios, TF+F achieves the highest exact and label accuracies, confirming the benefit of fusing local instances from the two domains.
Without fine-tuning, TF+F outperforms TF by 9.76 and 1.62 percentage points in exact and label accuracies, respectively, and outperforms F by 37.97 and 10.66 percentage points.
After few-shot fine-tuning, TF+F still improves over TF by 1.55 and 0.24 percentage points.
These results show that time-frequency instances provide the main discriminative evidence, while frequency instances are less effective alone but provide complementary information when fused with time-frequency instances.

\section{Conclusion}
This paper investigated compound interference recognition for LR-FHSS satellite IoT uplinks.
We formulated the task as a multi-instance multi-label learning problem and proposed a multi-domain instance fusion method that aggregates local instances from the time-frequency and frequency domains for bag-level multi-label recognition.
This formulation avoids treating each compound interference combination as an independent class and does not require instance-level annotations.

A dataset construction pipeline was developed under the US915 LR-FHSS configuration with shadowed-Rician fading and time-varying Doppler.
Experiments considered two practical receiver deployment scenarios, namely single-to-compound generalization and few-shot compound interference adaptation.
The proposed method achieved higher exact and label accuracies than representative baseline models in both scenarios.
These results show that localized interference evidence learned from non-interference and single interference samples can support recognition of unseen compound interference, and that a small number of labeled compound interference samples further improves complete multi-label prediction.

Further analyses show that time-frequency instances provide the main discriminative evidence, while frequency instances provide complementary information when fused with time-frequency instances.
Category-wise recall results indicate that reactive interference is the main source of single-to-compound generalization degradation, especially when RTI and RCHI coexist with other interference components.
Future work will validate the method with measured satellite IoT data and more diverse deployment conditions, including different LR-FHSS configurations, ISRs, and channel conditions.

\end{document}